\documentstyle[preprint,aps,pra,12pt]{revtex}

\tightenlines

\begin{document}
\preprint{CfA No. 4167}
\title{Nuclear spin--rotation interaction in the hydrogen molecular ion}

\author{J.~F. Babb}
\address{
Institute for Theoretical Atomic and Molecular Physics,\\
Harvard-Smithsonian Center for Astrophysics,\\
60 Garden Street, Cambridge, Massachusetts 02138
}

\maketitle

\begin{abstract}
The nuclear spin--rotation interaction in the hyperfine structure of
the hydrogen molecular ion is investigated.  The interaction constants
are determined and are found to differ in sign and magnitude compared
to another theory, but they are in agreement with some values derived
from experiment.
\end{abstract}

\narrowtext

Information on molecular structure~\cite{TowSch55}, nuclear
forces~\cite{CodRam71,ReiVai75}, fundamental
symmetries~\cite{KozLab95,SauWanHin95}, and even interstellar
molecules~\cite{SalSimWin94} can be gleaned from experimental and
theoretical studies of molecular hyperfine structure. The precision of
measurements of diatomic hyperfine transition frequencies is
continually improving because of developments such as ion trap/rf
spectroscopy~\cite{Jef69}, ion beam/laser
beam~\cite{WinRufLam76,CarMcNMon89}, and laser-radiofrequency
double-resonance~\cite{Chi92} methods.  Moreover, a recent
proposal~\cite{DoyFriKim95} for the trapping and cooling of
paramagnetic neutral molecules should offer---once spectroscopy has
been carried out---the ultimate in precision: natural linewidth
resolution of hyperfine frequencies.

For ${}^1\Sigma$ molecules, which have no net electronic spin, the
interaction energy of each nuclear magnetic moment with fields
generated by the motion of the other charged particles in the molecule
is important.  It has been studied in detail both experimentally and
theoretically for a number of such molecules (cf.  Ramsey~\cite{Ram56}
and Townes and Schawlow~\cite{TowSch55}).  The primary effect is the
nuclear spin--rotation interaction originating with the magnetic
fields generated by the rotating nuclei and orbiting electrons.  The
interaction constant can be related to the magnetic shielding
constant~\cite{Ram50d,Fly64}, which describes the effective magnetic
field at a nucleus in an external magnetic field and is the basis of
chemical shifts in NMR spectroscopy~\cite{Ram50d,Ram52e}.  It is also
a sensitive test of electronic wave function
calculations~\cite{Fly64,DocFre74,KomRycRay95}. Recently, nuclear
spin--rotation interaction has been interpreted as a (Berry or
geometric phase-like) manifestation of a nonabelian gauge potential in
molecular physics~\cite{SerSte95}.

Nuclear spin--rotation interaction is also present in ${}^2\Sigma$
molecules, which have a net electronic spin, but it causes a much
smaller energy in comparison to effects arising from the interactions
of the electron spin and thus it is usually not included in the
phenomenological spin Hamiltonians used for fitting measured hfs
transition frequencies.  Some exceptions are
$\mbox{H}_2{}^+$~\cite{Jef69}, $\mbox{N}_2{}^+$~\cite{BerManKur91},
and alkaline earth monofluorides~\cite{ChiGooRen81,SauWanHin95}, where
empirical values for nuclear spin--rotation interaction constants have
been obtained.

In this Letter, the nuclear spin--rotation interaction for the
hydrogen molecular ion $\mbox{H}_2{}^+$ is studied.  It is found that
for the ground vibrational state the interaction constant is
$-41$~kHz, nearly $\case{1}{2}$ the value for $\mbox{H}_2$,
$-113$~kHz~\cite{Ram56}, but differing from a previous theory in both
magnitude and sign.  For higher vibrational states there is agreement
with some empirical data.

Theories giving nuclear spin--rotation interaction constants for
diatomics with other than ${}^1\Sigma$ states exist, but they are
complex.  A simplification for $\mbox{H}_2{}^+$ is obtained by
utilizing the theory for $\mbox{H}_2$.  The isotropic magnetic
shielding constant $\sigma$, which describes the net magnetic field
seen at a nucleus in an external uniform applied magnetic field, is
related to the nuclear spin--rotation interaction constant.  The
shielding constant for nucleus ``a'' can be written
\begin{equation}
\label{sigma-HF-sub}
\sigma (R) = \sigma_{\rm L} (R) + \sigma_{\rm hf} (R) ,
\end{equation}
where, in atomic units, $R$ is the internuclear distance, and
\begin{equation}
\label{sigma-L}
\sigma_{\rm L}(R) = \case{1}{3}\alpha^2
           \langle 0| 1/r_a|0 \rangle
\end{equation}
is the ``Lamb''~\cite{Lam41,Ram56} or diamagnetic part, where ${\bf
r}_a$ joins the nucleus to the electron and $|0\rangle$ is the
${}^2\Sigma_g^+$ electronic wave function, and $\sigma_{\rm hf}(R)$ is
the high-frequency or paramagnetic part.  The designation
high-frequency~\cite{Van32} arises because the expression for
$\sigma_{\rm hf}$ involves highly-excited electronic
states~\cite{Ram56},
\begin{equation}
\label{sigma-HF}
\sigma_{\rm hf}(R)
   =   \frac{\alpha^2}{6}
        \sum_{i\neq 0} \left[
        \frac{\langle 0| {\bf L}_a|i \rangle \cdot
         \langle i| r_a^{-3}{\bf L}_a| 0 \rangle }
             {E_{0} (R) - E_{i} (R)} +
        \mbox{adj.}\right] ,
\end{equation}
where the symbol $\sum_{i\neq 0}$ represents an infinite
summation-integration over the intermediate electronic states with
wave functions $|i\rangle$, ``adj.'' indicates the Hermitian adjoint
of the preceding term, $E_0(R)$ and $E_i (R)$ are the respective
electronic energies and ${\bf L}_a$ is the orbital angular momentum of
the electron about the nucleus ``a'', for which $\sigma$ is being
evaluated.

The separation~(\ref{sigma-HF-sub}) into a diamagnetic and
paramagnetic part is not unique~\cite{Ram50d,Fly64}.  It is related to
the choice of origin for the electron orbital angular momentum and
other vectors, which in turn is related to the choice of a gauge
constant associated with the magnetic vector potential in the Coulomb
gauge~(cf.~\cite{Ram50d,Ram52e,Fly78}).  Expressions~(\ref{sigma-L})
and (\ref{sigma-HF}) above adopt the convention of the vector origins
at the nuclei.

Let ${\bf I}$ and ${\bf K}$ be, respectively, the total nuclear spin
and rotational angular momenta and define the nuclear spin--rotation
interaction constant $f$, a frequency, through the energy $hf{\bf
I}\cdot{\bf K}$ in the hfs Hamiltonian~\cite{IUPAC94,IUPAC-note},
where $h$ is Planck's constant.  Introducing,
\begin{equation}
\label{f-eq}
f(R) = f_1 (R) + f_2 (R) ,
\end{equation}
the major contributions~\cite{Ram56,small-note} to the energy are from
the interaction of each nuclear magnetic moment with the magnetic
field generated by the other rotating nucleus,
\begin{equation}
hf_1(R) = -\frac{4g_p\mu_N^2}{R^3}  ,
\end{equation}
and with the magnetic field generated by the orbiting
electron~\cite{Ram50d},
\begin{equation}
\label{f2}
hf_2(R) = -\frac{12g_p\mu_N^2}{\alpha^2R^2}  \sigma_{\rm hf} (R) ,
\end{equation}
where $R$ is the internuclear distance, and the dimensionless
quantities $\alpha$, $g_p\approx 5.586$, and $\sigma_{\rm hf}$ are,
respectively, the fs constant, the proton $g$-factor, and the
high-frequency component of the magnetic shielding constant as defined
in Eq.~(\ref{sigma-HF}). Atomic units are used throughout, except for
$f$, which is expressed in kHz.

Direct computation of $\sigma_{\rm hf} (R)$ is rather involved and
requires evaluation of a term in second-order perturbation theory.  It
is much easier to use (\ref{sigma-HF-sub}) because values of $\sigma$
are available and values of $\sigma_{\rm L}$ are easily calculable.
Moreover, because $\sigma$ is gauge-independent the irreducible
components of the (symmetric, second rank) shielding tensor calculated
with a different gauge origin by Hegstrom~\cite{Heg79} could be
utilized here to calculate $\sigma$~\cite{Sigma-note}.  Values of
$\sigma_{\rm L} (R)$ were calculated~\cite{sigma-L-calc} at various
internuclear distances $R$ and are given in Table~\ref{m-e}.  Then,
$\sigma_{\rm hf}$ was determined
using~(\ref{sigma-HF-sub}) and (\ref{sigma-L}) and values of $f$ were
determined using (\ref{f-eq}).  Values of $\sigma$ and $f$ are given
in Table~\ref{m-e}.

Jefferts measured hfs transition frequencies of the vibrational
$v=4$--$8$ states of the ${}^2\Sigma_g^+$ ground electronic state of
$\mbox{H}_2{}^+$.  For the rotational quantum number $K=1$ for each
$v$, he fit the transition frequencies to the Hamiltonian
\begin{equation}
\label{hfs-ham}
H_{\rm hfs} = b{\bf I}\cdot{\bf S} +
     c I_z S_z+ d {\bf S}\cdot{\bf K} + f{\bf I}\cdot{\bf K}
\end{equation}
and obtained values for the coupling constants $b$, $c$, $d$, and $f$,
where $\bf I$ and $\bf K$ have been defined, and $\bf S$ is the
electronic spin angular momentum vector, and his results for $f$ are
given in the Table~\ref{f-table}, col.~2.  The measured hfs transition
frequencies were refit to the Hamiltonian~(\ref{hfs-ham})
independently by Kalaghan~\cite{Kal72}, Menasian~\cite{Men73}, and
Varshalovich and Sannikov~\cite{VarSan93}, who all obtained mutually
consistent results for $f$ that had the opposite sign and somewhat
different magnitude than those of Jefferts. The values of~\cite{Kal72}
are given in Table~\ref{f-table}, col.~3.

McEachran, Cohen, and Veenstra~\cite{McEVeeCoh78} obtained a
theoretical expression for a nuclear spin--rotation interaction
constant by simply multiplying the first order {\em electronic\/}
spin-rotation constant $d_1$~\cite{DalPatSom60} by the ratio of the
electron mass and the nuclear reduced mass.  Their formula gives a
positive interaction constant.  Interestingly, it is independent of
the proton $g$-factor $g_p$, and thus does not contain the proton
magnetic moment.  Nevertheless, close agreement was obtained
with the empirical values of $f$ obtained by Jefferts.  The values for
$v=4$--8 of McEachran {\em et al.\/} and the $v=0$ value calculated
using their formula are given in column~5 of Table~\ref{f-table}.

Lundeen, Fu, and Hessels~\cite{FuHesLun92} extracted hfs constants for
$\mbox{H}_2{}^+$ from an analysis of their measured transition
energies for highly-excited Rydberg states of the hydrogen
molecule. Their value for $f$ for the $v=0$, $K=1$ state is given in
column 4 of Table~\ref{f-table}. It is consistent with zero.

In order to compare the present results with the above, the values of
$f$ in kHz---a partial listing is given in Table~\ref{m-e}---were
averaged over the vibrational-rotational wave functions for various
$v$ with $K=1$ calculated with the Born-Oppenheimer potential using
standard methods.  The results are given in the last column of
Table~\ref{f-table}.  Good agreement, including the sign, is obtained
with the empirical results of~\cite{Kal72,Men73,VarSan93}.

The expression~(\ref{f-eq}) can be obtained using more formal
arguments.  The first term in~(\ref{f-eq}), $f_1$, follows by
consideration of the interaction of a nuclear moment with the magnetic
field generated by the rotation of the other nucleus~\cite{FroFol52}.
The second  term in~(\ref{f-eq}),
$f_2$,  would be expected to occur in second order perturbation
theory through the electron orbital--rotation interaction
\begin{equation}
H_4 = -\frac{2}{M_pR^2} {\bf L}\cdot{\bf K}
\end{equation}
where ${\bf L}$ is the electronic angular momentum about the
center of nuclear mass and $M_p$ is the proton mass, and the electron
orbital--nuclear spin interaction
\begin{equation}
H_1' = H_{1a}' + H_{1b}' ,
\end{equation}
where
\begin{equation}
 H_{1a}'= 2 g_p\mu_{\rm B}\mu_N {\bf I}_a\cdot r_a^{-3}{\bf L}_a
\end{equation}
and
\begin{equation}
H_{1b}' = 2 g_p\mu_{\rm B}\mu_N   {\bf I}_b\cdot r_b^{-3}{\bf L}_b  ,
\end{equation}
where ${\bf I}_a$ and ${\bf I}_b$ are the nuclear spin angular
momenta, and where the terms $H_4$ and $H_1'$ were derived by
Dalgarno, Patterson, and Somerville~\cite{DalPatSom60} from a
non-relativistic reduction of the Dirac eq. for $\mbox{H}_2{}^+$.
Writing
\begin{equation}
H_4 =
 -\frac{1}{M_pR^2} ({\bf L}_a+{\bf L}_b)\cdot{\bf K}
  \equiv H_{4a} + H_{4b},
\end{equation}
where ${\bf L}_a$ is the electronic angular momentum about nucleus
$a$, and similarly for ${\bf L}_b$, the corresponding energy in
perturbation theory is, schematically, $T\equiv T_a + T_b$, where
\begin{equation}
\label{2-pert}
 T_a   = \sum_{i\neq 0} \left[\frac{\langle 0| H_{4a}|i\rangle
    \langle i| H_{1a}'|0\rangle }
 {E_0 (R) - E_i (R)}
  +  \mbox{adj.} \right] ,
\end{equation}
and similarly for $T_b$.  It follows that
\begin{equation}
\label{Ta}
T_a = \frac{-4g_p\mu_N^2}{R^2}
  \sum_{i\neq 0} \left[
   \frac{\langle 0|{\bf L}_a |i\rangle \cdot
         \langle i|r_a^{-3}{\bf L}_a |0\rangle}
         {E_0 (R) -E_i (R)}
  +  \mbox{adj.}  \right]
    {\bf I}_a\cdot{\bf K} ,
\end{equation}
and similarly for $T_b$.  Since ${\bf I} ={\bf I}_a +{\bf I}_b$ and
$T\equiv hf_2{\bf I}\cdot{\bf K}$, the desired result Eq.~(\ref{f2})
is obtained by substituting Eq.~(\ref{sigma-HF}) for $\sigma_{\rm hf}$
into $T_a$, Eq.~(\ref{Ta}), and into $T_b$, although a more careful
derivation is desirable.

Elaborate treatments of the Hamiltonians for diatomic molecules in
other than ${}^1\Sigma$ states including all angular momenta have
yielded nuclear spin--rotation interactions for various electronic
states~\cite{Fre66,Miz75,BroKaiKer78,BroVigLeh78,AdaAzuBar94}.  In
particular, terms similar to $f_1$ and $f_2$ were given by
Mizushima~\cite{Miz75} for a ${}^2\Sigma$ state, although they were
not related to Ramsey's theory.

The present approach may be generalizable to other ${}^2\Sigma$
molecules.  Hegstrom~\cite{Heg79} has shown for $\mbox{H}_2{}^+$ that
there is a relation between the second order {\em electron\/}
spin--rotation interaction constant $d_2$ and a second order part of
the shielding constant $\sigma$.  Eq.~(\ref{f2}) would thus imply that
there is a relation connecting $d_2$ and $f_2$.  Physically, such a
relation might be anticipated because the nuclear spin and electron
spin are both separately coupled to rotation through the excitation of
electronic angular momenta. Such a relation would make it possible to
estimate $f$, if $d$ were measured, using theoretical values of $f_1$
and $d_1$. Moreover, through the above arguments, the recent findings
that electron spin--rotation in a paramagnetic molecule and nuclear
spin--rotation in a diamagnetic molecule are each, separately,
describable using a nonabelian gauge potential method in molecular
physics~\cite{SteSer94,SerSte95} can probably be related.

The present paper demonstrates that there is a solid theoretical basis
for nuclear spin--rotation in a ${}^2\Sigma$ molecule.  A synthesis of
the theories for ${}^1\Sigma$ molecules and ${}^2\Sigma$ molecules has
yielded a simple expression for the interaction constant in
$\mbox{H}_2{}^+$. Using calculated properties of $\mbox{H}_2{}^+$
numerical values obtained for the interaction constant were found to
be in good agreement with empirical results. The sign of the effect is
found to be the same as for $\mbox{H}_2$, making $\mbox{H}_2{}^+$
another rare example of a diatomic with negative nuclear
spin--rotation interaction.

The author is grateful to Prof. K.~Burnett and Prof. A.~Dalgarno for
comments on the manuscript.  This work was supported in part by the
National Science Foundation through a grant for the Institute for
Theoretical Atomic and Molecular Physics at the Smithsonian
Astrophysical Observatory and Harvard University.


\clearpage

\begin{table}
\caption{\label{m-e}Electronic expectation values
of the shielding constants
$\sigma_{\rm L} (R)$ and $\sigma (R)$,
dimensionless, and values of $f(R)$, in kHz.}
\begin{center}
\begin{tabular}{lcrl}
\multicolumn{1}{c}{$R/a_0$} &
     \multicolumn{1}{c}{$\sigma_{\rm L} (R)\times 10^6$} &
     \multicolumn{1}{c}{$\sigma (R)\times 10^6$} &
     \multicolumn{1}{c}{$f(R)$} \\
\hline
  1.0    & 21.5183 & 17.6401 & $-$453.7 \\
  1.25   & 19.3436 & 15.3584 & $-$213.8 \\
  1.50   & 17.6258 & 13.6704 & $-$114.5 \\
  1.75   & 16.2484 & 12.3995 & $-$67.21 \\
  2.0    & 15.1295 & 11.4296 & $-$42.31 \\
  2.25   & 14.2116 & 10.6825 & $-$28.16 \\
  2.50   & 13.4530 & 10.1046 & $-$19.63 \\
  2.75   & 12.8232 &  9.6575 & $-$14.22 \\
  3.0    & 12.2991 &  9.3133 & $-$10.65 \\
  4.0    & 10.9538 &  8.6124 & $-$4.285 \\
  5.0    & 10.3478 &  8.4913 & $-$2.215 \\
\end{tabular}
\end{center}
\end{table}
\begin{table}
\caption{\label{f-table}Comparison of experimental and theoretical values
of the nuclear spin--rotation coupling constant for $\mbox{H}_2{}^+$
for various vibrational states $v$ with rotational quantum number
$K=1$, in kHz.  The column labeled ``Refit'' gives results obtained by
refitting the raw experimental transition frequencies
from~\protect\cite{Jef69}.  Numbers in parenthesis are quoted
experimental uncertainties.  }
\begin{center}
\begin{tabular}{cccccc}
\multicolumn{1}{c}{ } &
     \multicolumn{3}{c}{Empirical} &
     \multicolumn{2}{c}{Theory} \\
\cline{2-4} \cline{5-6}
\multicolumn{1}{c}{$v$} &
     \multicolumn{1}{c}{Expt.~\protect\cite{Jef69}} &
     \multicolumn{1}{c}{Refit\protect\tablenote{Values
are from Ref.~\protect\cite{Kal72} and  are identical with
those given in Refs.~\protect\cite{Men73} and \protect\cite{VarSan93},
except~\protect\cite{Men73} gives $-32 (1.5)$ for $v=5$.}} &
     \multicolumn{1}{c}{Expt.~\protect\cite{FuHesLun92}} &
     \multicolumn{1}{c}{\protect\cite{McEVeeCoh78}} &
     \multicolumn{1}{c}{Present\protect\tablenote{Values for
$v=1$, 2, and 3 are, respectively, $-40.8$, $-39.4$, and $-38.0$.}} \\
\hline\hline
0 &                &                & $-3(15)$    & 46    & $-41.7$    \\
\multicolumn{6}{c}{~}\\
4 & 38(1.5)        & $-34(1.5)$       &           & 38.8  & $-36.2$  \\
5 & 36(1.5)        & $-33(1.5)$       &           & 36.2  & $-34.6$  \\
6 & 34(1.5)        & $-31(1.5)$       &           & 33.6  & $-32.9$  \\
7 & 32(1.5)        & $-29(1.5)$       &           & 31.1  & $-31.0$  \\
8 & 30(1.5)        & $-27(1.5)$       &           & 28.6  & $-29.1$  \\
\end{tabular}
\end{center}
\end{table}


\begin{thebibliography}{10}

\bibitem{TowSch55}
C.~H. Townes and A.~L. Schawlow, {\em Microwave spectroscopy} (McGraw-Hill, New
  York, 1955), pp.\ 207--247.

\bibitem{CodRam71}
R.~F. Code and N.~F. Ramsey, Phys. Rev. A {\bf 4},  1945  (1971).

\bibitem{ReiVai75}
R.~V. Reid and M.~L. Vaida, Phys. Rev. Lett. {\bf 34},  1064  (1975).

\bibitem{KozLab95}
M.~G. Kozlov and L.~N. Labzowsky, J. Phys. B {\bf 28},  1933  (1995).

\bibitem{SauWanHin95}
B.~E. Sauer, J. Wang, and E.~A. Hinds, Phys. Rev. Lett. {\bf 74},  1554
  (1995).

\bibitem{SalSimWin94}
A.~H. Saleck, R. Simon, and G. Winnewisser, Ap. J. {\bf 436},  176  (1994).

\bibitem{Jef69}
K.~B. Jefferts, Phys. Rev. Lett. {\bf 23},  1476  (1969).

\bibitem{WinRufLam76}
W.~H. Wing, G.~A. Ruff, W.~E. Lamb~Jr., and J.~J. Spezeski, Phys. Rev. Lett.
  {\bf 36},  1488  (1976).

\bibitem{CarMcNMon89}
A. Carrington, I.~R. McNab, and C.~A. Montgomerie, J. Phys. B {\bf 22},  3551
  (1989).

\bibitem{Chi92}
W.~J. Childs, Phys. Rep. {\bf 211},  113  (1992).

\bibitem{DoyFriKim95}
J.~M. Doyle, B. Friedrich, J. Kim, and D. Patterson, Phys. Rev. A {\bf 52},  ??
   (1995), in press.

\bibitem{Ram56}
N.~F. Ramsey, {\em Molecular Beams} (Oxford University Press, Oxford, 1990),
  pp.\ 162--166, 208--213.

\bibitem{Ram50d}
N.~F. Ramsey, Phys. Rev. {\bf 78},  699  (1950).

\bibitem{Fly64}
W.~H. Flygare, J. Chem. Phys. {\bf 41},  793  (1964).

\bibitem{Ram52e}
N.~F. Ramsey, Phys. Rev. {\bf 86},  243  (1952).

\bibitem{DocFre74}
K. Kirby~Docken and R.~R. Freeman, J. Chem. Phys {\bf 61},  4217  (1974).

\bibitem{KomRycRay95}
J. Komasa, J. Rychlewski, and W.~T. Raynes, Chem. Phys. Lett. {\bf 236},  19
  (1995).

\bibitem{SerSte95}
Y.~A. Serebrennikov and U.~E. Steiner, Mol. Phys. {\bf 84},  627  (1995).

\bibitem{BerManKur91}
N. Berrah~Mansour, C. Kurtz, T. C. Steimle,
G. L. Goodman, L. Young, T.~J. Scholl, S.~D. Rosner, and R. A. Holt,
\newblock Phys. Rev. A {\bf 44}, 4418 (1991).


\bibitem{ChiGooRen81}
W.~J. Childs, L.~S. Goodman, and I. Renhorn, J. Mol. Spectrosc. {\bf 87},  522
  (1981).

\bibitem{Lam41}
W.~E. Lamb, Phys. Rev. {\bf 60},  817  (1941).

\bibitem{Van32}
J.~H. Van~Vleck, {\em The theory of electric and magnetic susceptibilities}
  (Oxford, London, 1932).

\bibitem{Fly78}
W.~H. Flygare, {\em Molecular structure and dynamics} (Prentice-Hall, Englewood
  Cliffs, NJ, 1978), pp.\ 392--400.

\bibitem{IUPAC94}
E. Hirota {\it et~al.}, J. Mol. Spectrosc. {\bf 168},  628  (1994).

\bibitem{IUPAC-note}
In Ramsey's notation $hc_W=-hf$ is the analogous coefficient. The
  IUPAC~\cite{IUPAC94} sign convention is adopted in the present paper.

\bibitem{small-note}
A much smaller effect due to terms arising from the acceleration of the nuclei
  is ignored in the application here to $\mbox{H}_2{}^+$ until experimental
  precision warrants its inclusion. See N.~F. Ramsey, Phys. Rev. {\bf 90}, 232
  (1953); R.~V. Reid~Jr. and A.~H.-M. Chu, Phys. Rev. A {\bf 9}, 609 (1974).

\bibitem{Heg79}
R.~A. Hegstrom, Phys. Rev. A {\bf 19},  17  (1979).

\bibitem{Sigma-note}
{In the notation of~\protect\cite{Heg79}, $\sigma = \case{1}{3} [ 2
  \sigma_\perp + \sigma_\|]$, where $\sigma_\perp = \sigma_\perp^{(1)} +
  \sigma_\perp^{(2)}$.}

\bibitem{sigma-L-calc}
The Born-Oppenheimer electronic wave functions of the ground state were
  calculated using the methods detailed in G. Hunter and H.~O. Pritchard, J.
  Chem. Phys. {\bf 46}, 2146 (1967), with $25$ terms in the polynomial
  expansion for each prolate spheroidal coordinate. The energies agreed with
  standard tabular values to at least 10 decimal places.

\bibitem{Kal72}
P.~M. Kalaghan, Ph.D. thesis, Harvard University, 1972.

\bibitem{Men73}
S.~C. Menasian, Ph.D. thesis, University of Washington, 1973.

\bibitem{VarSan93}
D.~A. Varshalovich and A.~V. Sannikov, Astron. Lett. {\bf 19},  290  (1993).

\bibitem{McEVeeCoh78}
R.~P. McEachran, C.~J. Veenstra, and M. Cohen, Chem. Phys. Lett. {\bf 59},  275
   (1978).

\bibitem{DalPatSom60}
A. Dalgarno, T.~N.~L. Patterson, and W.~B. Somerville, Proc. Roy. Soc. London
  {\bf 259A},  100  (1960).

\bibitem{FuHesLun92}
Z.~W. Fu, E.~A. Hessels, and S.~R. Lundeen, Phys. Rev. A {\bf 46},  R5313
  (1992).

\bibitem{FroFol52}
R.~A. Frosch and H.~M. Foley, Phys. Rev. {\bf 88},  1337  (1952).

\bibitem{Fre66}
K.~F. Freed, J. Chem. Phys {\bf 45},  4214  (1966).

\bibitem{Miz75}
M. Mizushima, {\em The theory of rotating diatomic molecules} (Wiley, New York,
  1975), p.~250, Eq.~(5-2-2).

\bibitem{BroKaiKer78}
J.~M. Brown, M. Kaise, C.~M.~L. Kerr, and D.~J. Milton, J. Mol. Spectrosc. {\bf
  36},  553  (1978).

\bibitem{BroVigLeh78}
M. Broyer, J. Vigu{\'e}, and J.~C. Lehmann, J. Phys. (Paris) {\bf 39},  591
  (1978).

\bibitem{AdaAzuBar94}
A.~G. Adam {\it et~al.}, J. Chem. Phys. {\bf 100},  6240  (1994).

\bibitem{SteSer94}
U.~E. Steiner and Y.~A. Serebrennikov, J. Chem. Phys. {\bf 100},  7503  (1994).

\end{thebibliography}
\end{document}